\documentclass[final,5p,times]{elsarticle}

\usepackage{graphicx}

\usepackage{amssymb}
\usepackage{amsthm}

\usepackage{lineno}

\journal{Physics Letters B}

\begin{document}

\begin{frontmatter}

\title{ Asymptotic characteristics of decay channels of light nuclei states in the ab initio
approach}

\author[label1]{D. M. Rodkin\corref{cor1}}
\address[label1]{Dukhov Research Institute for Automatics, 127055, Moscow, Russia}
\ead{rodkindm92@gmail.com}

\author[label1,label2]{Yu. M. Tchuvil'sky}
\address[label2]{Skobeltsyn Institute of Nuclear Physics, Lomonosov
Moscow State University, 119991 Moscow, Russia} 
\cortext[cor1]{I am corresponding author}

\begin{abstract}
A new convenient method for precise theoretical calculations of quantities of traditional theory of nuclear
reactions  such as  widths of resonances (including sub-threshold), and asymptotic
normalization coefficients is proposed. This method may be considered as a step on the road to full theoretical ab
initio description of light nuclei spectroscopic data. As an illustration of this method
the computational results for all relevant two-body channels for all known and some
theoretically predicted states of $^7$Li nucleus are shown.  Well-proven on a large amount of data 
Daejeon16 potential was used in the calculations. The most part of the results turn out to be in a
good agreement with the experimental data contained in spectroscopic tables.
\end{abstract}


\end{frontmatter}

Modern high-precision methods for describing light nuclei properties and reactions 
induced by light nuclei collisions are advancing nowadays. These methods are based on
new possibilities provided by modern high-performance supercomputers. An important role
among  the methods describing light nuclei structure belong to various ab initio methods,
such as different versions of No-Core Shell Model (NCSM) \cite{ncsm1, ncsm2, ncsm3, dyt1, dr},
Gamov Shell Model (GSM) \cite{gamov1, gamov2}, Green functions Monte Carlo method \cite{gfmc1,
gfmc2, gfmc3}, the Coupled Cluster Method \cite {ccm1} and Lattice Effective Field
Theory for Multi-nucleon Systems \cite{left1, left2, left3}. These methods are all based
on realistic NN-, NNN-, etc. potentials. These potentials could be derived from Chiral
Effective Field Theory \cite{dj16, machleidt1, machleidt2} or from nucleon
scattering data by the use of $J$-matrix inverse scattering method \cite{jisp1}. In the
current paper we use Daejeon16 potential \cite{dj16}. 

One of the most advanced ab initio methods for describing nuclear structure is No-Core
Shell Model. This model is based on solving A-nucleon Schr\"odinger equation using
realistic nucleon-nucleon potentials on the complete basis of totally antisymmetric
A-nucleon wave functions up to the maximal total number of
oscillator quanta N$^{max}_{tot}$. The size of this basis, for example, in widely used M-scheme reaches the value of about
10$^{10}$ in the case that a modern supercomputer is employed. This method 
could be used for calculations of the binding energies, sizes, beta-decay lifetimes and electromagnetic observables characterizing ground,
and excited states, as well as unstable resonances. 
 
It should be mentioned that the range of distances
where solutions of the Schr\"odinger equation are correctly described by NCSM  A-nucleon
wave functions (WFs) expands proportionally to $[N_{max}^{tot}]^{1/2}$. So this range is somewhat limited.

To  adopt microscopic approaches for description of nuclear reactions and the properties of nuclear resonance states
it is necessary to describe various cluster channels. They are under investigation within traditional
and modern models such as the  Resonating Group Model (RGM) \cite{rgm1, rgm2}, Algebraic Version of the 
RGM \cite{av1, av2}, Generator Coordinate Method 
\cite{hor,desc1}, Microscopic Cluster Model \cite{desc2,desc3}, THSR-approach
\cite{thsr}, Antisymmetrized Molecular Dynamics \cite{keh}. For the most part  these models are based on effective NN-potentials and used for calculations of specific highly clustered states.

Ab initio approaches focused on the discussed problem are also presented in the literature.
Among them the methods which combine NCSM and RGM namely No-Core Shell
Model / Resonating Group Model (NCSM/RGM) \cite{ncsmrg1} and No-Core Shell Model with
Continuum (NCSMC) \cite{ncsmc1, ncsmc2, ncsmc3, ncsmc4} seem to be the most versatile. Examples of 
a good description of the asymptotic  characteristics of decay channels of the spectra of $^7$Be
 and $^7$Li are presented in  \cite{ncsmc4}.  
As the NCSMC the Fermionic Molecular Dynamics (FMD) \cite{neff1,neff2,neff3} offers in fact an ab initio
approach focussed on the unified description of both bound states and continuum ones. 
Another approach -- Cluster Channel Orthogonal Functions Method (CCOFM) -- is also based on the employment of a basis 
combing NCSM and orthogonalized cluster-channel WFs \cite{our1, our2}. 
For the case of single-channel resonances some version of RGM \cite{kv} was put into
use. To describe the  resonance multi-neutron emission  the SS-HORSE method was
proposed \cite{sshorse1}. Some nuclear resonances can be also studied with the use
of mentioned above GSM. Anyway all these methods can be applied to a very limited number of nuclear states compare 
to the list of levels whose binding energies and electromagnetic properties are  described by NCSM. 

To make the mentioned above list of characteristics of nuclear states broader we proposed 
an approach \cite{our3} which is based on the calculations of the WFs of Hamiltonian eigenstates in the framework of conventional NCSM 
and projection of them onto the WFs of the cluster channels obtained in the framework of CCOFM \cite{our2}. 
In the current paper we present an advanced multi-channel version of the approach and the results of its application 
to the detail theoretical study of characteristics of $^7$Li nucleus spectrum. 

Let us demonstrate how translationally invariant A-nucleon WFs of arbitrary two-fragment decay channel with
separation A = A$_1$ + A$_2$ are built in CCOFM. The useful  feature of the procedure is that 
each function of this basis can be represented as a superposition of
Slater determinants (SDs) . To do that the technique of so-called cluster coefficients (CCs) is exploited. 

The oscillator-basis terms of the cluster channel $c_\kappa$ are expressed in the following form:
\begin{equation}
\Psi^{c_\kappa} _{{A\,},nl}  = \hat A\{\Psi^{\{k_1\}} _{A\,_1 } \Psi^{\{k_2\}} _{A\,_2 }
\varphi _{nl} (\vec \rho )\} _{J_cJM_JT} , \label{eq0}
\end{equation}
 where  $\hat A$ is the antisymmetrizer, $\Psi^{\{k_i\}}
_{A\,_i}$ is a translationally invariant internal WF of the fragment labelled by a set
of quantum numbers $\{k_i\}$; $\varphi _{nlm} (\vec \rho )$ is the WF of the relative
motion. The channel WF is labelled by the set of quantum numbers $c_\kappa$ which
includes $\{k_1\},\{k_2\},n,l,J_c,J,M_J,T$, where $J$ is the total momentum and $J_c$ is
the channel spin.  As it was mentioned above the function has to be represented
as a linear combination of the SDs containing the one-nucleon WFs
of the oscillator basis. For these purposes two-fragment WF in another representation
\begin{equation}
\Psi^{\{k_1,k_2\}} _{{A\,},nlm}  = \hat A\{\Psi^{\{k_1\}} _{A\,_1} \Psi^{\{k_2\}}
_{A\,_2} \varphi _{nlm} (\vec \rho )\} _{J_c,M_{J_c},M_JT}  \label{eq1}
\end{equation}
is multiplied by the function of the center of mass (CM) zero vibrations $\Phi _{000}
(\vec R)$. Then the transformation of WFs caused by changing from $\vec R,\vec \rho$ to
$\vec R_1 ,\vec R_2$ coordinates -- different-mass Talmi-Moshinsky transformation
 is performed \cite{talmi-moshinsky} and WF (\ref{eq1}) takes the form

\begin{equation}
\begin{array}{rcl}
\Phi _{000} (\vec R)\Psi^{\{k_1,k_2\}} _{{A\,},nlm} = \sum\limits_{N_i ,L_i ,M_i}
{\left\langle {{\begin{array}{*{20}c}
   {000}  \\
   {nlm}  \\
\end{array}  }}
 \mathrel{\left | {\vphantom {{\begin{array}{*{20}c}
   {000}  \\
   {nlm}  \\
\end{array}  } {\begin{array}{*{20}c}
   {N_1 ,L_1 ,M_1 }  \\
   {N_2 ,L_2 ,M_2 }  \\
\end{array}}}}
 \right. \kern-\nulldelimiterspace}
 {{\begin{array}{*{20}c}
   {N_1 ,L_1 ,M_1 }  \\
   {N_2 ,L_2 ,M_2 }  \\
\end{array}}} \right\rangle } \\[\bigskipamount]
\hat A\{ \Phi _{N_1 ,L_1 ,M_1 }^{A_1 } (\vec R_1 ) \Psi^{\{k_1\}}_{A\,_1 } \Phi _{N_2
,L_2 ,M_2 }^{A_2 } (\vec R_2 )\Psi^{\{k_2\}} _{A\,_2 } \} _{J_c,M_{J_c},M_JT} . \label{eq2}
\end{array}
\end{equation}

The main procedure of the method is to transform internal WFs corresponding to each
fragment multiplied by none-zero center of mass vibrations into a superposition of SDs

\begin{equation}
\Phi _{N_i ,L_i ,M_i }^{A_i } (\vec R_i )\Psi^{\{k_i\}}_{A\,_i }  = \sum\limits_k
{X_{N_i ,L_i ,M_i }^{A_i (k)} \Psi _{A\,_i (k)}^{SD} }. \label{eq3}
\end{equation}
Quantity $X_{N_i ,L_i ,M_i }^{A_i (k)}$ is called a cluster coefficient (CC). Technique of
these objects is presented in detail in \cite{nem}. There is a large number of methods
elaborated for the calculations of CCs. The most general scheme is based on the method
of the second quantization of the oscillator quanta. In this scheme the WF of the CM
motion is presented as

\begin{equation}
\Phi _{N_i ,L_i ,M_i }^{A_i } (\vec R_i ) = N_{N_i ,L_i } (\hat{ \vec \mu^{\dag}})^{N_i
- L_i } Y_{N_i ,L_i } (\hat{ \vec \mu^{\dag}})\Phi_{000}^{A_i} (\vec R_i ), \label{eq4}
\end{equation}
where  $\hat{ \vec \mu^{\dag}}$ is the creation operator of the oscillator quantum, and
$N_{N_i ,L_i }$ is the norm constant. Thus the CC turns out to be reduced to a matrix
element of the tensor operator expressed in terms of $ \hat{ \vec \mu^{\dag}}$:

\begin{equation}
\begin{array}{rcl}
 < \Psi _{A_i (k)}^{SD} |\Phi _{N_i ,L_i ,M_i }^{A_i } (\vec R_i ) \Psi^{\{k_i\}}_{A_i }  >  =
 N_{N_i ,L_i }
  \left\langle {\Psi _{A\,_i (k)}^{SD} }
 \right|  \\
 (\hat \mu ^\dag  )^{N_i  - L_i }
 Y_{N_i ,L_i } (\hat{ \vec \mu^{\dag}})
 \left| {\Phi _{000}^{A_i } (\vec R_i )\Psi^{\{k_i\}}_{A\,_i } } \right\rangle
\end{array}
\label{eq5}
\end{equation}

A conventional relationship between the translationally invariant and  an ordinary
shell-model WFs
\begin{equation}
\Psi _{A\,_i }^{shell} = \Psi^{\{k_i\}}_{A\,_i } * \Phi _{000}^{A_i } (\vec R_i )
\label{eq6}
\end{equation}
is used as a definition of the former one. The NCSM WFs of the fragments $\Psi _{A\,_i
}^{shell}$ are involved in the calculations.

The last step is to construct $\Psi^{c_\kappa} _{{A\,},n}$ basis wave functions for each
channel $c_\kappa$ (\ref{eq0}) from a basis of $\Psi^{\{k_1,k_2\}} _{{A\,},nlm}$ (\ref{eq1}) by the
use of ordinary Clebsh-Gordan coefficients.

It should be noted that WFs of cluster-channel terms (\ref{eq0}) of one and the same
channel $c_\kappa$ characterized by the pair of internal functions
$\Psi^{\{k_1\}}_{A_1}$, $\Psi^{\{k_2\}}_{A_2}$  and extra quantum numbers
$l,J_c,J,M_J,T$ are non-orthogonal. Creation of orthonormalized basis functions of
channel $c_\kappa$ is performed by the diagonalization of the exchange kernel

$ ||N_{nn'} || =  < \Psi^{c_\kappa} _{{A\,},n'}|\Psi^{c_\kappa} _{{A\,},n} > = $
\begin{equation}
< \Psi^{\{k_1\}} _{A_1} \,\Psi^{\{k_2\}} _{A_2} \,\varphi _{nl} (\rho )
|\hat A^2 |\Psi^{\{k_1\}} _{A_1} \,\Psi^{\{k_2\}} _{A_2} \,\varphi _{n'l} (\rho ) > . \label{eq7}
\end{equation}

The eigenvalues and eigenvectors of this exchange kernel are given by the expressions:

\begin{equation}
\varepsilon _{\kappa,k}  =  < \hat A\{ \Psi^{\{k_1\}} _{A_1} \,\Psi^{\{k_2\}} _{A_2} \,f_l^k (\rho
) \} |\hat 1|\hat A\{ \Psi^{\{k_1\}} _{A_1} \,\Psi^{\{k_2\}}_{A_2} \} \,f_l^k (\rho
)  > ; \label{eq8}
\end{equation}

\begin{equation}
f_l^k (\rho ) = \sum\limits_n {B_{nl}^k \varphi _{nl} (\rho )}. \label{eq9}
\end{equation}

As a result, the WF of the orthonormalized basis channel basis $c_\kappa$ is
represented in the form

\begin{equation}
\Psi^{SD,c_\kappa} _{{A\,},kl}=\varepsilon^{-1/2} _{\kappa,k}|\hat A\{ \Psi^{\{k_1\}}
_{A_1} \,\Psi^{\{k_2\}}_{A_2} \,f_l^k (\rho ) \}  >, \label{eq10}
\end{equation}

The cluster form factor (CFF) is a projection of the function of an initial A-nucleon state $\Psi_{A}$ 
 onto the WF of a particular channel $c_\kappa$. It  describes the relative motion of subsystems and  has the form

$$\Phi^{c_\kappa}_A(\rho)=\sum\limits_k\varepsilon^{-1/2} _{\kappa,k}<\Psi _{A}|\hat A\{ \Psi^{\{k_1\}} _{A_1}
\,\Psi^{\{k_2\}}_{A_2} \,f_l^k (\rho ) \}>f_l^k (\rho )$$
\begin{equation}
=\sum\limits_k\varepsilon^{-1/2} _{\kappa,k}\sum\limits_{n, n'}B_{nl}^k B_{n'l}^{k}
C_{AA_1A_2}^{n'l}\varphi _{nl} (\rho ) \label{eq11}
\end{equation}
where

$C_{AA_1A_2}^{nl} = < \hat A\{\Psi^{\{k_1\}} _{A_1} \Psi^{\{k_2\}} _{A_2} \varphi _{nl}
(\rho ) \} | \Psi _{A} >=$
\begin{equation}
<\Psi^{SD,c_\kappa} _{A, nl}|\Phi _{000} (R)| \Psi _{A} >=<\Psi^{SD,c_\kappa} _{A,
nl}|\Psi^{SM} _{A} >. \label{eq12}
\end{equation}
is called the spectroscopic amplitude. A number of very diverse methods of its
calculation depending on the masses of the initial nuclei and fragments were described
in \cite{nem,sa1, sa2, sa3}. The amplitude of the CFF is determined as

\begin{equation}
K_{AA_1A_2}^{nl}=\sum\limits_{k, n'}\varepsilon^{-1/2} _{\kappa,k}B_{nl}^k B_{n'l}^{k}
C_{AA_1A_2}^{n'l} .\label{eq13}
\end{equation}
The spectroscopic factor of this channel $c_\kappa$ has the form

$$S_l^{c_\kappa}  = \int {|\Phi^{c_\kappa}_A(\rho)|^2 } \rho ^2 d\rho =$$
\begin{equation}
\sum\limits_k {\varepsilon _k^{ - 1} \sum\limits_{nn'} {C_{AA_1A_2}^{nl} }
C_{AA_1A_2}^{n'l} B_{nl}^k } \;B_{n'l}^k . \label{eq14}
\end{equation}

The definitions of the CFF (\ref{eq12}) and spectroscopic factor (\ref{eq14}) are completely
equivalent to those proposed in \cite{newsf1} (the so-called “new” spectroscopic
factor and CFF). In contrast to the traditional definition, the new
spectroscopic factor characterizes the total contribution of orthonormalized cluster
components to the solution of the Schr\"odinger equation for A nucleons. Reasons for the
necessity of its use to describe decays and reactions can be found in \cite{newsf2,
newsf3}. In \cite{vt1, vt2}, it was demonstrated that the correct definition eliminates
a sharp contradiction between theoretical calculations of the cross sections for
reactions of knockout and transfer of $\alpha$ clusters and experimental data.

Compared to the spectroscopic factor the CFF is a more informative characteristic because it
allows its matching with the asymptotic WF of the relative motion in the range of validity of shell-model solution
and thus determines the amplitude of the WF in the asymptotic region. 
 
It is convenient to use the CFF for computing the widths of resonances and the
 asymptotic normalization coefficients (ANCs) of bound states, which in turn are used to
calculate the cross sections of resonant and peripheral reactions respectively. The CFF in its new
definition allows matching with the asymptotic WF at relatively small
distances, where the nuclear interaction is negligibly weak, but exchange effects
generated by the antisymmetry of the total channel WF are still not negligibly small.
This property is very important for dealing with NCSM WFs.

In this approach, a direct matching procedure is applied to calculate the ANCs of bound states and the widths of narrow resonances. For such resonances or, more precisely, for those of them whose small width is due to a
low penetrability of the potential barrier, we used a very compact procedure proposed in
\cite{matchingpoint}. This procedure is applicable because  for such resonances there is  sufficiently
wide range of distances in which the nuclear attraction is already switched off and at the same time 
the potential barrier is high enough.
At any inner point $\rho_{in}$ of this area, the relationship between the regular and irregular solutions 
of the two-body Schr\"odinger equation in the WKB approximation has the form

\begin{equation}
F_l (\rho_{in} )/ G_l (\rho_{in} ) = P_l (\rho_{in} ) \ll 1\label{eq15}
\end{equation}
where $P_l (\rho_{in} )$ is the penetrability of the part of the potential barrier which
is located between the point $\rho_{in}$ and the outer turning point. The smallness of
this penetrability is the condition of applicability of the approximation, where the
contribution of the regular solution can be neglected.
To determine the position of the matching point of the CFF and irregular WF in this
range, we use the condition of equality of the logarithmic derivatives

\begin{equation}
\frac{d\Phi _A^{c_\kappa  } (\rho)/d\rho}{\Phi _A^{c_\kappa  } (\rho )}= \frac{dG_l
(\rho)/d\rho}{{G_l (\rho )}}, \label{eq16}
\end{equation}
which determines the matching point $\rho_{m}$; therefore, the decay width is given
by the expression

\begin{equation}
\Gamma  = \frac{{\hbar ^2 }}{{\mu k}}\left [ \frac{{\Phi _A^{c_\kappa  } (\rho
_{m})}}{{G_l (\rho _{m})}}\right ]^2. \label{eq17}
\end{equation}

The matching of logarithmic derivatives is also used to determine the ANCs $A^{c_\kappa}$. 
This value  is defined as a ratio of the
CFF and the Whittaker function $W_{-\eta,l+1/2}(2k\rho)$ in the range of
distances where the nuclear interaction is already weak \cite{anc1, anc2}

\begin{equation}
A^{c_\kappa}= \rho_{m} \Phi _A^{c_\kappa} (\rho_{m} )/W_{-\eta,l+1/2}(2k
\rho_{m}). \label{eq18}
\end{equation}

In several spectroscopic data tables \cite{ndt} the decay
widths of sub-threshold resonances are given. To establish a link with such data we
calculate the decay widths of such a type by the use of the formulation given in \cite{subthreshold}:

\begin{equation}
\Gamma_{subth}(E)= \frac{\hbar^2}{\mu_{ab}} k \rho_{m} P_l (E)
\frac{W^2_{-\eta,l+1/2}(2k_0 \rho_{m})}{\rho_{m}} |A^{c_\kappa}|^2 \label{eq19}
\end{equation}
where $k_0 = \sqrt{2 \mu_{ab} c^2 E_{state}} / (\hbar c)$. 

To make the list of the properties  of the states of a certain nucleus complete  large-width
resonances are considered too. If the resonance is wide and so the penetrability $P_l
(\rho_{in} )$ (\ref{eq15}) is not small the width of this resonance is calculated using the
 simple version of the conventional R-matrix theory:

\begin{equation}
\Gamma  = \frac{\hbar^2}{\mu k_0} (F^2_l ( \rho_{m} ) + G^2_l ( \rho_{m} ))^{-1}
(\Phi _A^{c_\kappa} (\rho_{m} ))^2. \label{eq20}
\end{equation}
Naturally the use of this version leads to reduction in accuracy of calculation results, but it shoul be noted
that the accuracy of the the data, concerning large-width resonances, extracted from various experiments is also 
very limited. Thus, together with calculations using NCSM model our method allows one to calculate
simultaneously nearly all properties of ground and excited states of light nuclei. 

The critical point of the approach is a correct representation of the form of the CFF at distances at 
which, first, the nuclear interaction is negligible and, second, the "memory" of the exchange effects 
remains exclusively in the exchange kernel matrix $ ||N_{nn'} ||$. 
In this work the proposed approach is tested by computing of the asymptotic characteristics of 
all relevant two-body channels for all known and some
theoretically predicted states of $^7$Li nucleus together with the spectrum of this nuclide. 
 
An important point is that the Daejeon16 potential \cite{dj16} is used as a model of NN-interaction. 
It is built using the N3LO limitation for of Chiral Effective Field Theory \cite{ceft1} softened by Similarity
Renormalization Group (SRG) transformation \cite{srg1}. This potential is designed to calculate 
all kinds of characteristics of nuclei with the masses $A \leq 16.$ starting from one and the same
set of the parameters of NN-interaction. It was tested in the framework of broad calculations of the
total binding energies, excitation energies, radii,  moments of nuclear states and the electromagnetic reduced probabilities. 
These tests demonstrated that such characteristics are, in general, reproduced well. To calculate
the WFs of the $^{7}$Li nucleus, and its subsystems within the NCSM, we use the Bigstick
shell-model code \cite{bigstick}, which is convenient for multiprocessor computer clusters. The
basis cutoff parameter  $N^{max}_{tot} = 15$ was reached for the computation of the
energies and WFs of the $^7$Li nucleus states. The optimal oscillator parameter $\hbar
\omega$ for the calculations with the realistic NN-potential Daejeon16 turns out to be 20.0 MeV.

\begin{flushleft}
\begin{table}
\caption{Amplitudes of the CFF $K_{AA_1A_2}^{nl=1}$ of all “nodal” quantum numbers for the
ground state
of $^7$Li and $^4$He+$^3$H channel in various bases used to calculate WFs of initial nuclei and subsystems.}
\begin{tabular*}{0.5\textwidth}{c|lllll}
\hline\hline\noalign{\smallskip}
$N_{tot}^{max}$ &$N_{cl}^{max}$ & n=1 &n=3 & n=5 &n=7\\
\hline\noalign{\smallskip}
13& 0 & 0.0 & 0.745 & -0.419 & 0.262 \\
      & 2 & 0.007 &0.752 &-0.435 &0.271  \\
      & 4 & 0.092 & 0.753 &-0.440  &0.0  \\
\hline\noalign{\smallskip}
15& 0 & 0.0       &0.729  &-0.421  & 0.273  \\
      & 2 & 0.006 &0.736  &-0.436  & 0.282  \\
      & 4 & 0.090 &0.737  &-0.443  &0.288  \\
\noalign{\smallskip}\hline\hline
\end{tabular*}
\end{table}
\end{flushleft}

According to equations (\ref{eq17},\ref{eq18}, \ref{eq19}, \ref{eq20}) the ANCs 
and decay widths of nuclear states depend on the value of
channel form factor at the matching point. The coefficients (amplitudes)
$K^{nl=1}_{AA_1A_2}$ of its oscillator expansion for the ground state of $^7$Li and
$^4$He + $^3$H channel are summarized in Tab. 1. The data presented in this table
indicate that the dominant amplitudes of the CFF come down to their real
values already for the basis cutoff parameter $N^{max}_{tot}= 13$ and the choice of
the cutoff parameter $N_{cl}^{max}$ of the bases of the cluster WFs weakly affects these
quantities. The measure of the convergence can be defined as the total standard deviation of
the amplitudes of the CFF in different calculation variants. If one
excludes the case $N^{max}_{tot}= 13$ and $N_{cl}^{max}= 4$, where the dimension of the
basis is insufficient to calculate the amplitude which is characterized by the “nodal” quantum number n =
7 , this deviation becomes less than 1$\%$. So the channel form factor doesn't depend
noticeably on the accuracy of the subsystem description and the choice of nearly all combinations of
subsystems cut off parameters is adequate.

\begin{figure}[htp]
\includegraphics[scale=0.45]{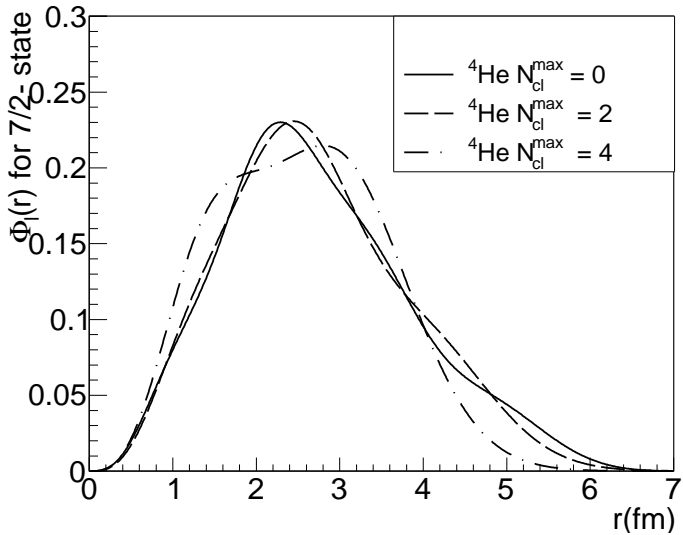}
\caption{ CFFs $\Phi_{l}(r)$ for 7/2$^{-}_1$ state of $^{7}$Li for $^4$He+$^3$H channel as
function of relative distance for different values of cutoff parameter $N_{cl}^{max}$ of the cluster WF.
\label{fig1}}
\end{figure}

As another demonstration of the fact that CFF depends only slightly on
the choice of the cutoff parameter $N_{cl}^{max}$, the CFFs for
7/2$^{-}_1$ state and $^4$He+$^3$H channel are shown in Fig. \ref{fig1}. 

\begin{figure}[htp]
\includegraphics[scale=0.45]{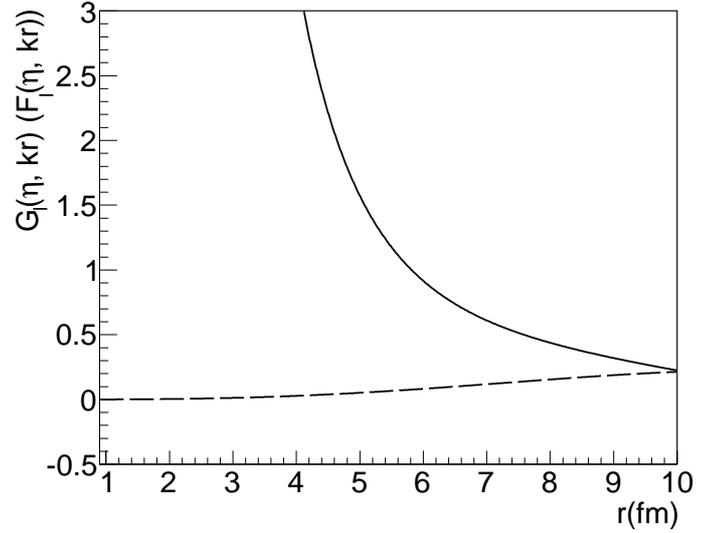}  
\caption{ Asymptotic regular (dashed line) and irregular (solid line) solutions for 7/2$^{-}_1$ state of
$^{7}$Li for $^4$He+$^3$H channel.
\label{fig2}}
\end{figure}

For all low-lying resonances and bound states the channel form factors are matched with asymptotic
solutions. As an example these asymptotic solutions are shown for 7/2$^{-}_1$ state and
$^4$He+$^3$H channel in Fig. \ref{fig2}. For this state the condition $F_l (\rho_{in}
)\ll G_l (\rho_{in} )$ is exactly satisfied in fact. So the contribution of
the regular solution can be neglected and one may determine the position of the matching
point taking into account the irregular solution only. 

\begin{figure}[htp]
\includegraphics[scale=0.45]{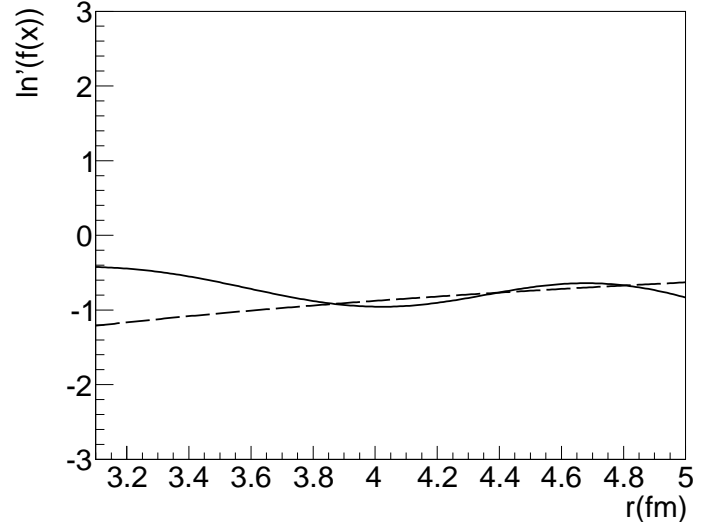} 
\caption{ Logarithmic derivatives of CFF $\Phi_{l}(r)$ (solid line) and irregular asymptotic
solution (dashed line) for 7/2$^{-}_1$ state of $^{7}$Li for $^4$He+$^3$H channel. \label{fig3}}
\end{figure}

Bearing in a mind mentioned above problems appearing in the NCSM calculations of WFs at large 
inter-fragment distances a reproducibility of the form of asymptotic WF by the CFF 
in this region turns out to be a key question. Fig.3 demonstrates a high quality of the reproducibility
in the discussed example. The ratios of the CFFs and the asymptotic WFs change weakly in the  region of
matching point (see bellow Fig. \ref{fig5}). Due to that the the result of the 
matching procedure turn out to be tolerant to occasional 
deviation of the matching point in rather wide range of distances. Moreover this circumstance allows one to
overcome difficulties which could occur in the case that a synthesis of the using projection procedure and the
 microscopic R-matrix theory \cite{rmatr} would be required. 

As it is shown in Tabs. 2 -- 4 the total binding energies of the lower levels of the $^7$Li nucleus obtained in our
calculations  reproduce the measured values with accuracy conventionally considered satisfactory and are in good agreement with the NCSM calculations, which we received by us
through personal communication with the authors of this potential. The energies of highly-excited states are
reproduced with a lower accuracy especially for states with relatively small spin or positive parity. 
The insufficient dimension of the basis employed and, possibly, some features of the potential used are the causes of that.
At the same time, the performed calculations demonstrate that the asymptotic
characteristics of the channels under study are highly sensitive to the fragmentation energy. In
particular, the ANC of 3/2$^-_1$ state varies from 3.44  to 3.30 fm$^{-1/2}$ (by 3\%) at
replacement of the experimental energy E$_{exp.}$=-2.467 MeV   by the computed value  E$_{th.}$=-2.529 MeV, i.e.
at the change in the energy by 62 keV. This difference for the 1/2$^-_1$ state, the change is equal to 0.29 MeV in this example (it is a rare case when the excitation energy is poorly reproduced in our and all other calculations  using 
great bases hence originated by features of the Daejeon16 potential itself),  is 24\%. The decay widths are even more sensitive to the fragmentation energy. Indeed, in the case of 7/2$^-_1$ state a very small only 23 keV difference in resonance energy (in the case of replacement of the experimental energy 2.195 by the calculated one -- 2.172 MeV) leads to 5\% difference in decay width (from 65 keV to 62.2 keV). It is clear enough that modern ab initio computations cannot reproduce the fragmentation energy of the decay channels with the precision that required for the discussed purposes, therefore the calculations which are based on the experimental values of the decay energies are preferable. We follow that choice. Our calculation show that the coordinates of the matching point turn out to be rather stable and have  a little impact on the results.

To complete the presentation of the methodical aspects of the proposed approach it is appropriate to compare 
them with the features of the approach designed in Ref.  \cite{ncsmc4}, where $^7$Li and $^7$Be 
asymptotic properties were studied with the use of NCSMC model. As it was mentioned above, the NCSMC
 model is a well self-consistent ab initio approach to unified description of nuclear 
structure and reactions. It is focussed on the description of compound states of nuclei together 
with their real or virtual, nucleon or light cluster decay channels. A realistic NN-potential softened via the 
SRG transformation is used in the discussed paper. 
The value of the parameter of this transformation $\lambda_{SRG}$ = 2.15 fm$^{-1}$ has been
chosen as the best that reproduces the experimental separation energies of nucleons and clusters in 
$^7$Be and $^7$Li nuclei. The NCSMC model  allows one to compute ANCs of bound states and partial 
decay widths of resonances, both narrow and wide. At present the capability of this approach are limited
by one-channel problem and the mass of the light fragment A$\leq3$. The approach presented in the current paper
looks more simple and the mass of the light fragment is not a strong limitation. The asymptotic characteristics of all 
two-body decay channels of a certain state can be calculated simultaneously. The well-tested on the large basis of 
nuclear data Daejeon16 NN-potential is used in the approach. A possibility to involve 
into consideration experimental values of decay energies makes the approach more flexible. In a lot of examples
this increases the quality of description of the spectroscopic data. That is why we believe that the developed approach 
extends the possibilities of ab initio methods to describe characteristics of various of nuclear states and
their decay channels and makes the description more accurate in many cases. 
What about the correlation of the numerical results obtained by the discussed approaches one may conclude that 
the results presented in Ref. \cite{ncsmc4} and the results shown in the current paper bellow being different in 
particular examples can be generally qualified as good ones.  

The results of the calculations of asymptotic characteristics of all relevant virtual and real 
decay channels ($^4$He + $^3$H, $^6$Li +n, $^6$He + p, 
$^6$Li(E$^*$ = 3.562 MeV, T = 1) + n) for all known $^7$Li and some theoretically 
predicted states  are summarized in Tabs. 2 -- 4.

\begin{center}
\begin{table*}
\caption{Asymptotic characteristics of $^7$Li for $^4$He+$^3$H and $^6$Li+n channels for
bound negative parity states and high-excited positive parity state.}
\begin{tabular*}{0.97\textwidth}{ c c c c c c c c }
\hline\hline\noalign{\smallskip}
\begin{tabular}{c}J$^{\pi}$,\\T  \end{tabular} & \begin{tabular}{c} E$_{exp.}$(E$_{th.}$) \\ MeV \end{tabular} & \begin{tabular}{c} E$^{exp.}_{\alpha + t}$ \\ (E$^{th.}_{\alpha + t}$) \\ MeV \end{tabular} & \begin{tabular}{c} $\Gamma_{\alpha}$(ANC) \\ \cite{exp7Li} \end{tabular} &
\begin{tabular}{c} $\Gamma_{\alpha}$(ANC) \\ th. \end{tabular} & \begin{tabular}{c} E$^{exp.}_n$ \\ (E$^{th.}_n$) \\ MeV \end{tabular} & \begin{tabular}{c} $\Gamma_n$(ANC) \\ \cite{ndt, gn1} \end{tabular}  & \begin{tabular}{c} $\Gamma_n$(ANC) \\ th. \end{tabular}\\

\hline\noalign{\smallskip}  \begin{tabular}{c}3/2$^{-}$, \\ 1/2 \end{tabular} & \begin{tabular}{c} 39.245 \\ (39.110) \end{tabular}  & \begin{tabular}{c} -2.467 \\ (-2.529) \end{tabular}  &  \begin{tabular}{c} 3.57 $\pm$
0.15 \\ fm$^{-1/2}$ \end{tabular}  & \begin{tabular}{c} 3.44  \\ fm$^{-1/2}$ \end{tabular} & \begin{tabular}{c} -7.25 \\ (-7.639) \end{tabular} &
\begin{tabular}{c}
th.results \cite{gn1} \\
$l=1, J_{n}=1/2$ \\
1.652 fm$^{-1/2}$\\
$l=1, J_{n}=3/2$ \\
1.890 fm$^{-1/2}$ \\
\end{tabular}
&
\begin{tabular}{c}
$l=1, J_{n}=1/2$ \\
-1.618 fm$^{-1/2}$ \\
$l=1, J_{n}=3/2$ \\
1.317 fm$^{-1/2}$\\
\end{tabular}
 \\
\hline\noalign{\smallskip}

 \begin{tabular}{c} 1/2$^{-}$, \\ 1/2 \end{tabular} & \begin{tabular}{c}  38.768\\ (38.279) \end{tabular} & \begin{tabular}{c} -1.98 \\ (-1.69) \end{tabular}  & \begin{tabular}{c} 3.0 $\pm$ 0.15 \\ fm$^{-1/2}$ \end{tabular}  & \begin{tabular}{c} 2.95 \\
fm$^{-1/2}$ \end{tabular} & \begin{tabular}{c} -6.77 \\ (-6.808) \end{tabular} &
\begin{tabular}{c}
th.results \cite{gn1} \\
$l=1, J_{n}=1/2$ \\
-0.540 fm$^{-1/2}$\\
$l=1, J_{n}=3/2$ \\
-2.540 fm$^{-1/2}$
\end{tabular}
&
 \begin{tabular}{c}
$l=1, J_{n}=1/2$ \\
-0.531 fm$^{-1/2}$\\
$l=1, J_{n}=3/2$ \\
1.979 fm$^{-1/2}$
\end{tabular}
 \\
\hline\noalign{\smallskip}

 \begin{tabular}{c} $1/2^{+}$, \\ 1/2 \end{tabular} & \begin{tabular}{c} 32.804 \\ (28.921) \end{tabular} & \begin{tabular}{c} 3.984 \\ (7.66) \end{tabular} & \begin{tabular}{c} 3.15 \\ MeV \end{tabular}  & \begin{tabular}{c} 7.4 
\\ MeV \end{tabular} & \begin{tabular}{c} - 0.81 \\(2.55) 
\end{tabular} &
\begin{tabular}{c}
$\Gamma (E=1eV)$ \\
0.29 keV \\
\end{tabular}
&
\begin{tabular}{c}
$l=0, S=1/2$ \\
$\Gamma (E=1eV)$ \\
0.54 keV \\
\end{tabular}
 \\
\hline\hline\noalign{\smallskip}
\end{tabular*}
\end{table*}
\end{center}

\begin{figure}[htp]
\includegraphics[scale=0.45]{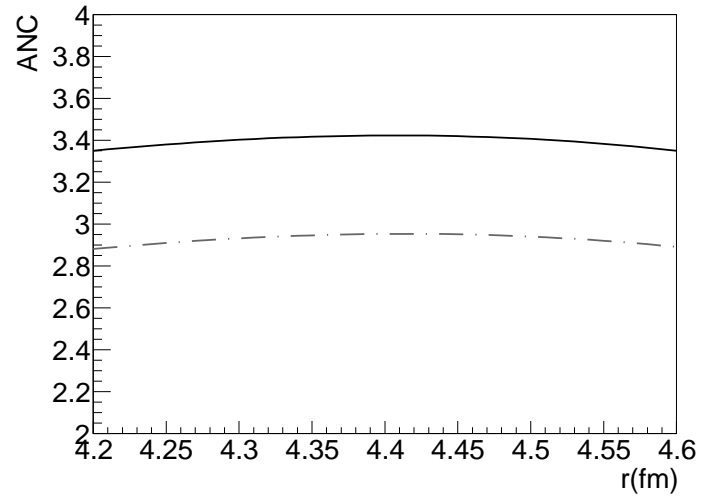}
\caption{Asymptotic normalization coefficient in the region of matching point for the
3/2$^-$ (solid line) and 1/2$^-$ (dash-dotted line) states. \label{fig4}}
\end{figure}

\begin{center}
\begin{table*}
\caption{Asymptotic characteristics of $^7$Li for $^4$He+$^3$H and $^6$Li+n channels for
high-excited negative parity states.}
\begin{tabular*}{0.95\textwidth}{c c c c c c c c}
\hline\hline\noalign{\smallskip}
J$^{\pi},T$ & \begin{tabular}{c} E$_{exp.}$(E$_{th.}$) \\ MeV \end{tabular} &  \begin{tabular}{c} E$^{exp.}_{\alpha + t}$ \\ (E$^{th.}_{\alpha + t}$) \\ MeV \end{tabular} & \begin{tabular}{c} $\Gamma_{\alpha}$(ANC) \\ \cite{exp7Li, toi} \end{tabular}
& \begin{tabular}{c} $\Gamma_{\alpha}$(ANC) \\ th. \end{tabular} & \begin{tabular}{c} E$^{exp.}_n$ \\ (E$^{th.}_n$) \\ MeV \end{tabular} & \begin{tabular}{c} $\Gamma_n$(ANC) \\ \cite{ndt, gn1} \end{tabular} & \begin{tabular}{c} $\Gamma_n$(ANC) \\ th. \end{tabular} \\
\hline\noalign{\smallskip}

7/2$^{-},1/2$ & \begin{tabular}{c} 34.593 \\ (34.409) \end{tabular} & \begin{tabular}{c}  2.195 \\ (2.172) \end{tabular} & \begin{tabular}{c} 69 \\ keV \end{tabular}  & \begin{tabular}{c}  65 \\ keV \end{tabular} & \begin{tabular}{c}  -2.59 \\ (-2.938) \end{tabular} &
-----
&
\begin{tabular}{c}
$l=3, S=1/2$ \\
0.013 fm$^{-1/2}$ \\
\end{tabular}
 \\
\hline\noalign{\smallskip}

5/2$^{-},1/2$ & \begin{tabular}{c} 32.641 \\ (31.610) \end{tabular} & \begin{tabular}{c}  4.147 \\ (4.971) \end{tabular} & \begin{tabular}{c} tot.: 918 \\ keV \end{tabular}  & \begin{tabular}{c} 564 \\ keV \end{tabular} & \begin{tabular}{c}  -0.65 \\ (-0.139) \end{tabular} &
-----
&
\begin{tabular}{c}
$l=1, S=3/2$ \\
0.199 fm$^{-1/2}$ \\
\end{tabular}
 \\
\hline\noalign{\smallskip}

5/2$^{-},1/2$ & \begin{tabular}{c} 31.791 \\ (30.816) \end{tabular} & \begin{tabular}{c} 4.997 \\ (5.765) \end{tabular} & \begin{tabular}{c} 22.9 \\ keV \end{tabular}  & \begin{tabular}{c} 797 \\ keV \end{tabular} & \begin{tabular}{c}  0.2 \\ (0.655) \end{tabular} & \begin{tabular}{c} 57.7 \\ keV \end{tabular} &
\begin{tabular}{c}
$l=1, S=3/2$ \\
53 keV \\
\end{tabular}
 \\

\hline\noalign{\smallskip}

3/2$^{-},1/2$ & \begin{tabular}{c} 30.495 \\(28.175) \end{tabular} & \begin{tabular}{c} 6.293 \\ (8.406) \end{tabular} & \begin{tabular}{c} tot.: 4.7 \\ MeV \end{tabular}  & \begin{tabular}{c} 873 \\ keV \end{tabular} & \begin{tabular}{c} 1.5 \\ (3.296) \end{tabular} & \begin{tabular}{c} 0.867 \\ MeV \end{tabular} &
\begin{tabular}{c}
$l=1, S=1/2$ \\
88 keV \\
$l=1, S=3/2$ \\
1.0 MeV \\
\end{tabular}
 \\
\hline\noalign{\smallskip}

1/2$^{-},1/2$ & \begin{tabular}{c} 30.155 \\(27.280) \end{tabular} & \begin{tabular}{c} 6.633 \\ (9.301) \end{tabular} & \begin{tabular}{c} tot.: 2.7 \\ MeV \end{tabular} & \begin{tabular}{c} 282 \\ keV \end{tabular} & \begin{tabular}{c} 1.84 \\ (4.191) \end{tabular} &
-----
&
\begin{tabular}{c}
$l=1, S=1/2$ \\
230 keV \\
$l=1, S=3/2$ \\
1.0 MeV \\
\end{tabular}
 \\
\hline\noalign{\smallskip}

7/2$^{-},1/2$ & \begin{tabular}{c} 29.675 \\ (28.489) \end{tabular} & \begin{tabular}{c} 7.133 \\ (8.092) \end{tabular} & \begin{tabular}{c} tot.: 437 \\ keV \end{tabular}  & \begin{tabular}{c} 453 \\ keV  \end{tabular}& \begin{tabular}{c} 2.32 \\ (2.982) \end{tabular} &
-----
&
\begin{tabular}{c}
$l=3, S=1/2$ \\
0.72 keV \\
\end{tabular}
 \\
\hline\noalign{\smallskip}

3/2$^{-},1/2$ & \begin{tabular}{c}  ----  \\ (27.047) \end{tabular} &\begin{tabular}{c}   ---- \\ (9.60) \end{tabular} & ----  & \begin{tabular}{c}  1.25 \\ MeV \end{tabular} & \begin{tabular}{c} ---- \\
(4.424) \end{tabular} &
-----
&
\begin{tabular}{c}
$l=1, S=1/2$ \\
 785 keV \\
\end{tabular}
 \\

\hline\hline\noalign{\smallskip}
\end{tabular*}
\end{table*}
\end{center}

\begin{center}
\begin{table*}
\caption{Asymptotic characteristics of high-excited states of $^7$Li for $^6$He(gr.)+p
and $^6$Li(E*=3.562 MeV, T=1)+n channels.}
\begin{tabular*}{0.88\textwidth}{c c c c c c c c c }
\hline\hline\noalign{\smallskip}
J$^{\pi},T$ & \begin{tabular}{c} E$_{exp.}$(E$_{th.}$) \\ MeV \end{tabular} & \begin{tabular}{c}  E$^{exp.}_{He + p}$ \\ (E$^{th.}_{He + p}$) \\ MeV \end{tabular} & \begin{tabular}{c} $\Gamma_{tot}$\\ \cite{exp7Li} \end{tabular} & SF$_{He + p}$ &
\begin{tabular}{c} $\Gamma_{He + p}$ \\ (ANC) th. \end{tabular} &  \begin{tabular}{c} E$^{exp.}_{Li + n}$ \\ (E$^{th.}_{Li + n}$) \\ MeV \end{tabular} & SF$_{Li+n}$ &  \begin{tabular}{c} $\Gamma_{Li+n}$ \\ (ANC) \end{tabular} \\

\hline\noalign{\smallskip} 1/2$^{-},1/2$ & \begin{tabular}{c} 30.155 \\ (27.280) \end{tabular} & \begin{tabular}{c} -0.885 \\ (1.538) \end{tabular}  & ---  & 0.1465 &
\begin{tabular}{c} 0.343 \\ fm$^{-1/2}$ \end{tabular} & \begin{tabular}{c} -1.727 \\ (0.638) \end{tabular} & 0.0453 & \begin{tabular}{c} 0.259 \\ fm$^{-1/2}$ \end{tabular}
 \\
\hline\noalign{\smallskip} 3/2$^{-},3/2$ & \begin{tabular}{c} 28.005 \\ (27.247) \end{tabular} & \begin{tabular}{c} 1.265 \\ (1.571) \end{tabular}  & \begin{tabular}{c} 260 \\ keV \end{tabular}  &
0.1638 & \begin{tabular}{c} 111 \\ keV \end{tabular} & \begin{tabular}{c} 0.433 \\ (0.671) \end{tabular} & 0.3770 & \begin{tabular}{c} 117 \\ keV \end{tabular}
\\
\hline\hline\noalign{\smallskip}
\end{tabular*}
\end{table*}
\end{center}

\begin{figure}[htp]
\includegraphics[scale=0.45]{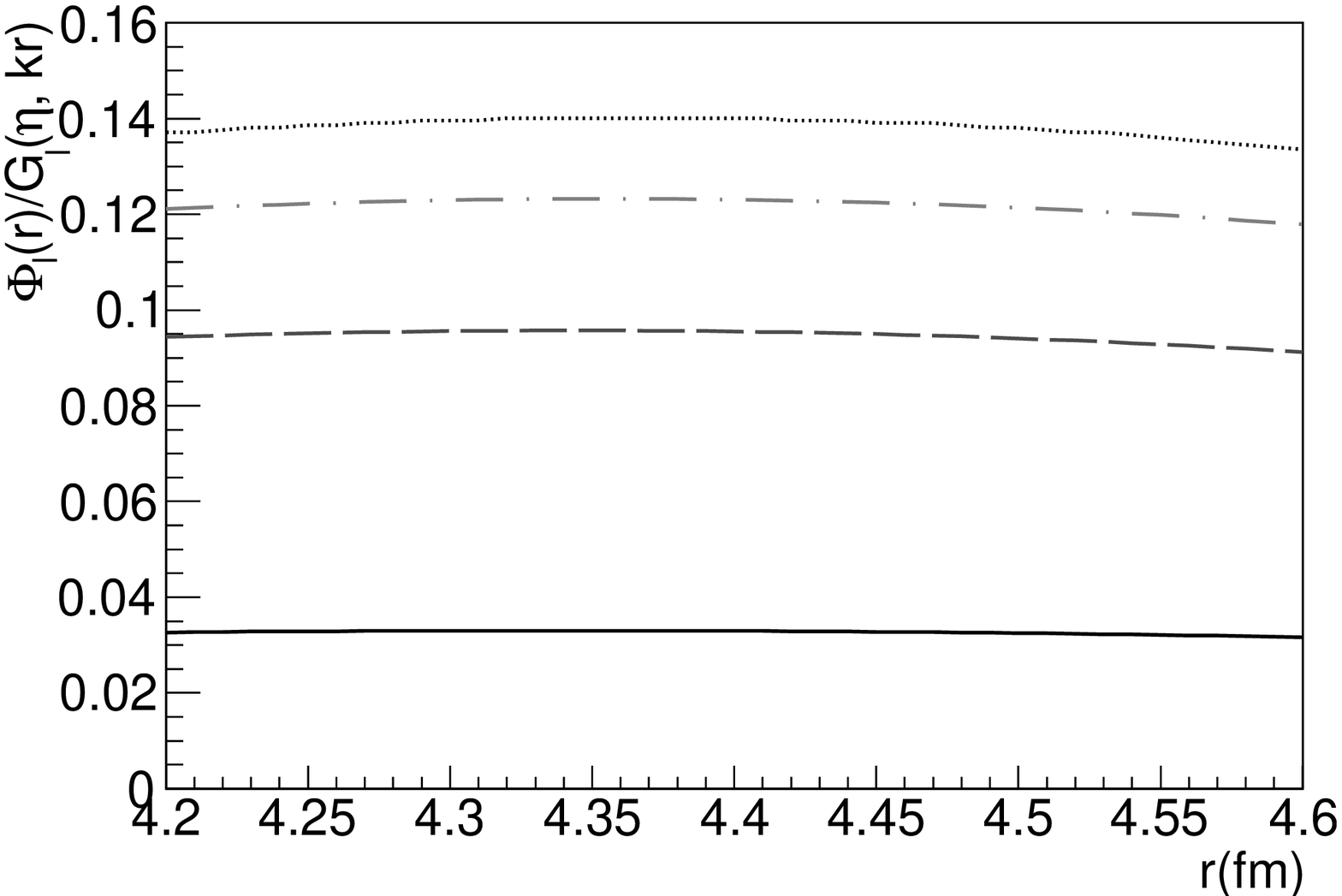} 
\caption{ The ratio of the CFF and the irregular Coulomb function ($\Phi_l
(r)/ G_l (\eta, kr)$ in the region of the matching point for the  7/2$^-_1$ (solid line),
5/2$_1^-$ (dashed line), and 5/2$_2^-$ (dashed-dotted line) states and the ratio of the
CFF $\Phi_{l}(r)$ and $({G^2_l (\eta, kr) + F^2_l (\eta, kr)})^{1/2}$ (dotted line) for high-exited 7/2$^-_2$
state.\label{fig5}}
\end{figure}

The matching points for 3/2$^-$ and 1/2$^-$ states for $^4$He + $^3$H channel
almost coincide. It is important that ratio (\ref{eq18}) near their matching point
hardly changes for both states, as seen in Fig. \ref{fig4}. This confirms once more the
stability of the procedure used to calculate the ANCs.
The results are also in a good agreement with the values extracted from experiments (see
Tab. 2).

The ratios of CFFs and irregular asymptotic solutions for 7/2$^-_1$,
5/2$^-_1$ and 5/2$^-_2$ states are also stable (see Fig. \ref{fig5}). They are located  in the sub-barrier
region and satisfy the condition $F_l (\eta, kr) \ll G_l (\eta, kr)$ and, thus, the
condition of equality of the logarithmic derivatives of CFFs and
irregular solutions is valid. 

In the case of highly-excited state 7/2$^-_2$ the inequality (\ref{eq15})
is not satisfied, so the conventional  $R$-matrix
theory (\ref{eq20}) is used. The ratio of CFF of the $^4$He + $^3$H channel and 
$(G^2_l (\eta, kr) + F^2_l (\eta, kr))^{1/2}$ according to Fig. \ref{fig5} is stable. This example shows
that our method remains applicable even for high-excited states.

In our previous paper \cite{our3} it was shown that  7/2$^-_1$ state is  strongly
clustered. Its calculated decay width is in a very good agreement with the experimental
value as it is shown in Tab. 3. In Ref. \cite{exp7Li}, this value is presented as an
alternative to a commonly accepted value 93 keV \cite{toi}. Taking into account a high
stability of the calculations performed, we consider the result as a serious argument in
favor of the value shown in \cite{exp7Li}. In contrast to the successful description of
the asymptotic characteristics of three lowest levels, the width of the  5/2$_2^-$ level
calculated with the Daejeon16 potential is strongly inconsistent with the experimental
value though the aggregate decay width of two  5/2$_2^-$ levels is reproduced more or less
well. We guess that this discrepancy is due to the fact that these states are
overlapped. Indeed, the energy distance ($\sim$800 keV) between them is smaller than the
aggregate decay width observed in the experiments. The small distance makes the WFs of
the overlapping states to be sensitive to a small variation of the parameters of the
used potential. In the discussed case the large overestimation of 5/2$_2^-$ state decay
width is possibly determined by the contribution of the components with the spin $S$ =
1/2. The excess of these components obtained in the calculations for this state and,
correspondingly, their deficit in the 5/2$_1^-$ state indicate some features of the
description of spin-dependent  forces by the potential in use. It is clear that any
calculation can hardly reproduce  the statistical weights of small components
with a proper accuracy; meanwhile, the presence of such components with the spin
$S$ = 1/2 determines the asymptotic characteristics of the $^4$He + $^3$H channel for
this state.  It is interesting to note that problem of reproducing the energy gap
between 3/2$^-_1$ and 1/2$^-_1$ states in the computations performed with the use of 
Daejeon16 potential is obviously associated with the spin-dependent  part of it too.  Perhaps there
are other effects originated by overlapping of  5/2$_{1,2}^-$ resonance states.

The decay widths of $^4$He + $^3$H channel are calculated also for other highly-excited
states such as 3/2$_2^-$, 3/2$_3^-$ and 1/2$_2^-$. The results do not
contradict much with known data, especially bearing in a mind that the widths of 
this highly-excited states couldn't be  extracted with a high accuracy from experimental data.

The asymptotic characteristics of the discussed states are calculated also for other
decay channels. For $^6$Li +n channel our results demonstrate a good agreement with
known decay widths for 5/2$_2^-$ and even for exotic 1/2$^+$ state. In the case of
5/2$_2^-$ it is even more surprising since as it is shown above the decay width for
$^4$He + $^3$H is overestimated. From this it follows  that possible defects in the used NN-potential 
does not strongly affect the WF as a whole but only some number of its
components. It should be mentioned that the presented method is applicable even for
calculations of the asymptotic characteristics of neutron $s$-wave channels as it is shown by the
example of 1/2$^+$ state. For other states not much information is known. So  the
results just presented are the most complete for neutron decay channel of $^7$Li
nucleus.

The discussed method is also used for channels containing exotic subsystems (isobar
analogues $^6$He (ground) and $^6$Li (E$^*$ = 3.562 MeV, T = 1)). The results related to only two states
1/2$^-_{2}$ and 3/2 (T=3/2) are  presented. It is because only they have significant values of the 
spectroscopic factors for these channels and only they are located in the near-threshold
region. Solely the total width of 3/2 (T=3/2) state ($\simeq  260$ keV) is known from the
experimental data and the calculated value agree with it perfectly since the aggregate
two-channel decay width of this state  is 228 keV. For sub-threshold state
1/2$^-_{2}$ neither experimental nor other  theoretical data concerning  these two
channels are known. Nevertheless it is obvious that eventually the method showed its
high reliability and wide applicability in this example ones more.

The main results of this work are the following.

\noindent
I. Within the framework of outgrows of the Cluster Channel Orthogonal Functions Method an ab initio scheme for computing the asymptotic characteristics of nuclear states, namely the asymptotic normalization coefficients of bound states and partial decay  widths of resonances is designed. This method is founded on the conventional NCSM and the projection procedure of WFs of the model onto the cluster channel functions. At the same time not only NCSM WFs but other ones obtained in an ab initio approach and representable in the form of shell-model expansion can be exploited in the approach.  

\noindent
II. This scheme allows one to obtain simultaneously the asymptotic characteristics of arbitrary
two-fragment real and virtual decay channels of all known states of light nuclei.
 
\noindent
III. The asymptotic characteristics of the bound and resonant states of $^7$Li are
calculated for all two-body decay channels of interest. Good agreement with the
experimental data is achieved not only for the dominating decay channels, but also for
channels with exotic subsystems. The results of the computations demonstrate prospects 
of the method to obtain experimentally
unknown ANCs of bound states and partial decay widths of resonant states of light nuclei.

\noindent
IV. It is demonstrated that the approach may be promising as an additional test on
NN-interaction used as an input.

\noindent
V. The performed calculations of the asymptotic characteristics of all low-lying states of
light nuclei will allow, in the short term, to achieve a keen insight into delicate
microscopic effects when calculating the cross sections of resonant and peripherical 
nuclear reactions induced by collisions of light nuclei.

\noindent
VI. The authors believe that the current work may turn out to be a step on the way to
creation of the theoretical nuclear spectroscopy which allows one to reproduce and predict 
all observables characterizing light nuclei.

\bibliography{sample}

\end{document}